\documentclass[showpacs,twocolumn,showkeys,preprintnumbers,amsmath,amssymb]{revtex4}

\usepackage{graphicx}
\usepackage{dcolumn}
\usepackage{bm}
\usepackage{amsmath,amssymb,amsfonts,amssymb,amsthm}

\begin{document}
\title{Method for arbitrary phase transformation by a slab
based on transformation optics and the principle of equal optical
path}

\author{Yougang Ke}
\author{Weixing Shu}\thanks{wxshu@hnu.edu.cn}
\author{Hailu Luo}
\author{Shuangchun Wen}\thanks{scwen@hnu.edu.cn}
\author{Dianyuan Fan}

\affiliation{Key Laboratory for Micro-/Nano-Optoelectronic Devices
of Ministry of Education, College of Information Science and
Engineering, Hunan University, Changsha 410082, China}


\begin{abstract}
The optical path lengths travelled by rays across a wavefront
essentially determine the resulting phase front irrespective of
the shape of a medium according to the principle of equal optical
path. Thereupon we propose a method for the transformation between
two arbitrary wavefronts by a slab, i.e. the profile of the
spatial separation between the two wavefronts is taken to be
transformed to a plane surface. Interestingly, for the mutual
conversion between planar and curved wavefronts, the method reduce
to an inverse transformation method in which it is the reversed
shape of the desired wavefront that is converted to a planar one.
As an application, three kinds of phase transformation are
realized and it is found that the transformation on phase is able
to realize some important properties such as phase reversal or
compensation, focusing, and expanding or compressing beams, which
are further confirmed by numerical simulations.  The slab can be
applied to realizing compact electromagnetic devices for which the
values of the refractive index or the permittivity and
permeability can be high or low, positive or negative, or near
zero, depending on the choice of coordinate transformations.
\end{abstract}

\pacs{42.15.Dp, 41.20.Jb, 42.79.-e, 02.40.-k}
\keywords{phase transformation, wavefront, optical path
length, transformation optics}
\maketitle

\section{Introduction}

In the last few years, transformation optics
\cite{Pendry2006,Leonhardt2006}, based on the form-invariance of
Maxwell's equations under coordinate transformations, has been
attracting more and more attention. According to the equivalence
of space geometry and material in routing light, it provides a
robust method to control electromagnetic waves by materials
usually implemented by metamaterials \cite{Schuring2006a}. This
has led to remarkably successful applications to designing
electromagnetic devices, which may be mainly classified into four
classes. The first is to control the path of light ray, such as
invisibility cloaks \cite{Schuring2006b,Zhang2011PRL,Chen2011NC},
light concentrator/absorber \cite{Rahm2008a,Narimanov2009}, beam
expanders/compressors \cite{Xu2008a,Garcia-Meca2011}, beam shifter
\cite{Rahm2008c,Gallina2010}, and beam bends
\cite{Rahm2008b,Liu2010}. The second is to manipulate the
polarization \cite{Kwon2008b} and the third is to control the
amplitude \cite{Luo2008PIER,Kundtz2008,Li2009}. While the last is
to mold the phase with notable examples as the field rotators
\cite{Chen2009} and phase transformers
\cite{Jiang2008b,Yu2011,Ma2008a}.

There exists a limitation on the wave through most of the above
devices yet. That is, the exit wave would return
\cite{Schuring2006b} or be parallel \cite{Rahm2008c} to the
incident wave direction, or be perpendicular to the exit interface
\cite{Rahm2008b} because the wave vector or phase front has not
been changed by transformations.  Recently, phase transformers in
the last class achieved the direction of exit wave different from
the incoming one, but at the cost of irregular profiles due to the
direct transformation used. It is natural to ask if a device can
be designed independent of the geometrical shape, such as a slab,
to manipulate the exit wavefront. As well known the planar
structure is preferred in practical applications, especially as
plug-and-play devices, being compact enough as well as free of
aberrations \cite{Kundtz2010}. Indeed certain planar designs have
been used to realize directive emission with high performance by
transforming cylindrical waves into plane ones
\cite{Zhang2008,Kwon2008a,Lin2008a,Tichit2011,Kong2007,Luo2011}.
However a general theory for phase transformation through planar
configurations has not been proposed yet.

In this work we introduce a new coordinate transformation method
from the fundamental concepts of wavefront and optical path length
(OPL) in ray optics to design a slab to manipulate the phase.
Through the slab the conversion between two arbitrary wavefronts
can be achieved. The slab can be realized by materials for which
the refractive index or the permittivity and permeability can be
high or low, positive or negative, or near zero, depending on the
choice of transformations. The results can be applied to realizing
compact electromagnetic devices, such as wave deflector, flat
lens, phase compensator and beam expander or compressor. In
addition, the method takes a straightforward form and can
completely avoid possible singular points in constitutional
parameters.

This work is organized as follows. In Sec.~\ref{pformulation} we
derive from the concepts of wavefront and OPL the coordinate
transformation method related to planar wavefronts, i.e. the
inverse transformation method, then extend it to a general form in
Sec.~\ref{cformulation}. In Sec.~\ref{application} we realize
three kinds of phase transformation as an application of the
approach and discuss the influence of the choice of coordinate
transformation on material parameters. Discussion and conclusion
are made in the last section.

\section{The basic formulation}\label{formulation}

Let us begin with the concept of wavefront and OPL. As well known
the wavefront is the surface of constant phase and the phase
differences of any two corresponding points on two sequential
wavefronts are identical \cite{Hecht2002}. The optical path
lengths (OPLs) involved are also equal, stated by the principle of
equal optical path \cite{Born1999}. In accordance with Fermat's
principle, the rays across initial wavefront traverse individual
paths so that the OPLs or the times of travel are extrema, whether
through conventional optical components \cite{Saleh2007} or newly
developed transformation media. If these OPLs are distinct, they
arrive the exit surface at different moments and the associated
phases are varied. Thereby the resultant wavefront is distorted
according to the principle of equal optical path. So, strictly
speaking, it is the OPLs travelled rather than the shape of medium
that fundamentally determine the resulting phase front
\cite{Born1999,Saleh2007}. Theoretically, therefore, one phase
transformation can be realized by materials of various profiles as
long as the needed OPLs, i.e. phase differences, are satisfied. On
the other hand, the transformation between wavefronts corresponds
to one specific coordinate transformation in a mathematical
perspective \cite{Chen2009,Jiang2008b,Yu2011,Ma2008a,Kundtz2010,
Zhang2008,Kwon2008a,Lin2008a,Tichit2011,Kong2007,Luo2011}. Thus it
is entirely feasible to construct material with a slab
configuration to transform wavefront by transformation optics.
These form the basis for our method.

\subsection{The inverse transformation method for
the transformation associated with planar
wavefronts}\label{pformulation}

\begin{figure}\centering
\includegraphics[width=4cm]{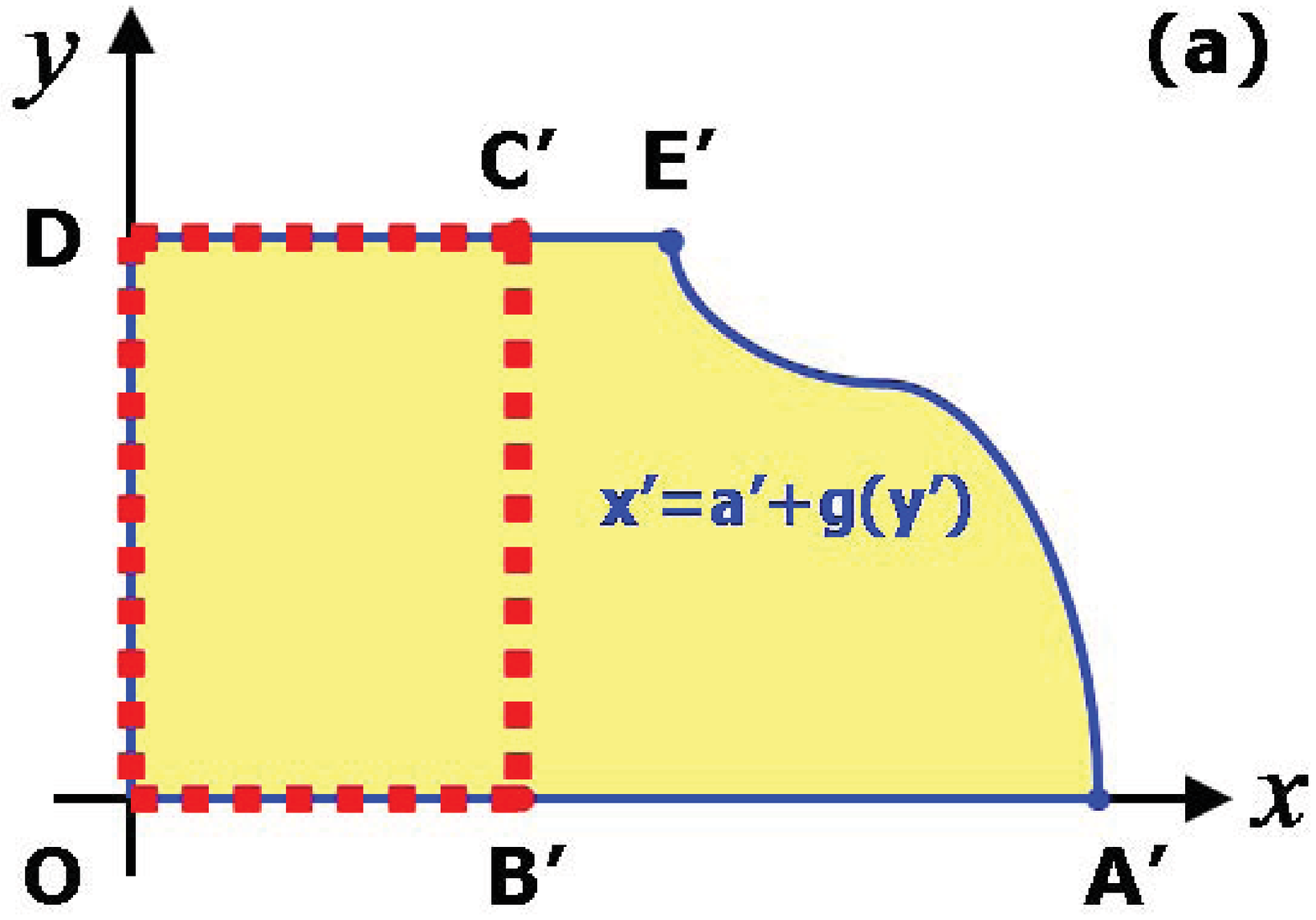}
\includegraphics[width=4cm]{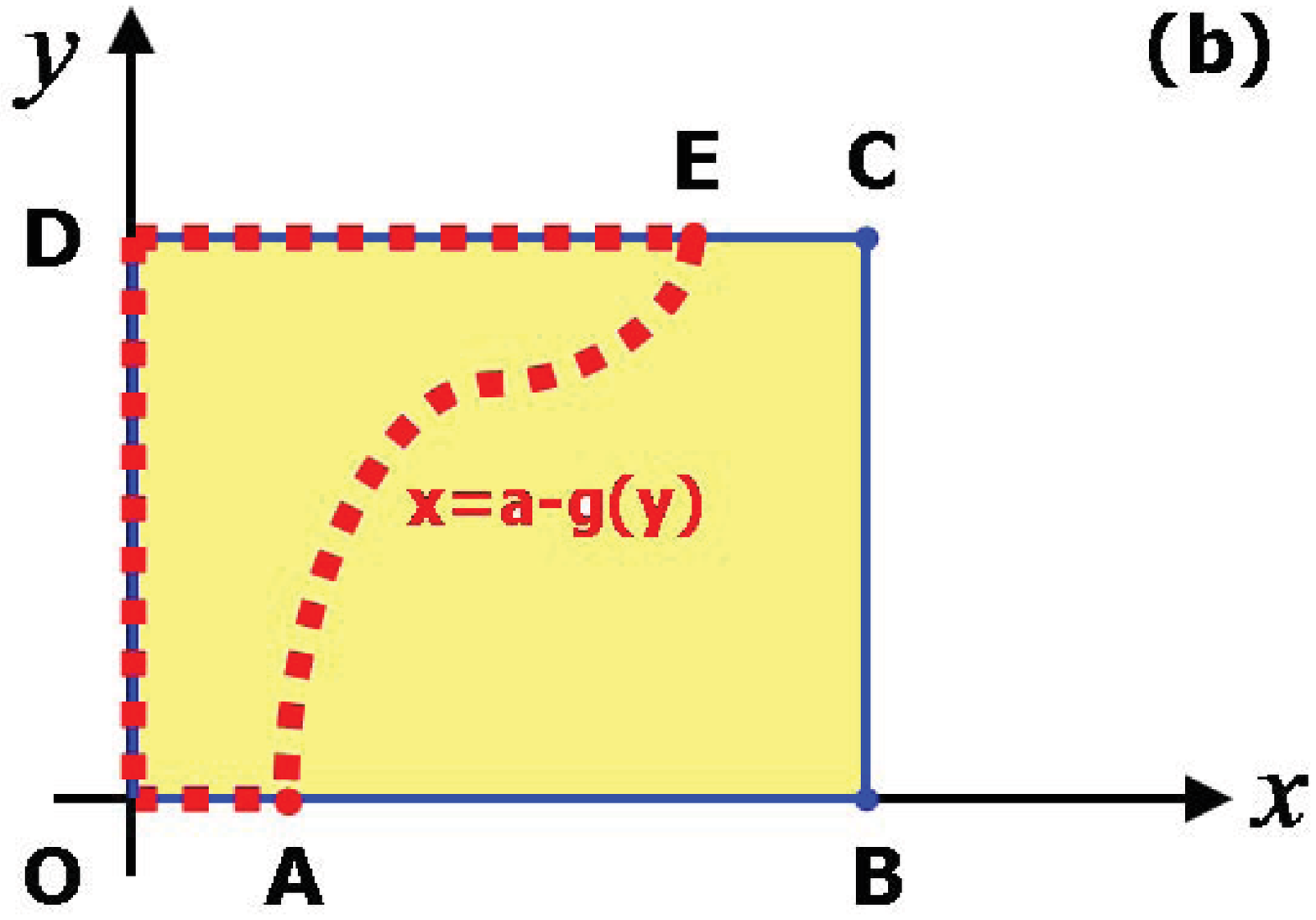}
\caption{\label{Schem} Schematic diagram of the space
transformation to convert a planar phase front into a curved one
by (a) direct transformation method in  literature and (b) inverse
transformation method adopted in the present work.  $AE$ is a
reversed version of $A'E'$. The dotted regions are the virtual
spaces while the real-lined regions indicate the physical spaces.
$OA'=a', OB'=b', OA=a, OB=b$, and $OD=d$.}
\end{figure}

For simplicity we consider the two-dimensional wave problem and
first establish the framework of transformation to convert a
planar wave front into any desired phase front. To achieve that
the usual approach in literature is to transform a plane surface
to the desired wave shape immediately, i.e. from $B'C'$ to $A'E'$
as shown in Fig.~\ref{Schem}(a), which may be called direct
transformation method (DTM) \cite{Jiang2008b,Yu2011,Ma2008a}.
Therein the virtual region ($OB'C'D$, enclosed by dotted lines) is
transformed into the physical one ($OA'E'D$ in yellow). Such a DTM
is intuitive enough but results in designs with irregular shapes,
$OA'E'D$. Instead, we adopt an inverse transformation method
(ITM): \emph{the reversed shape of the desired wavefront is
converted to a planar one}, i.e., from $AE$ to $BC$, whereby the
design is a slab, $OBCD$, as shown in Fig.~\ref{Schem}(b). Here it
is the virtual region $OAED$ that is transformed into the flat
physical one $OBCD$.

In the transformation of any point from $B'C'$ to $A'E'$, the
distance travelled is $g(y')+a'-b'$, where $B'C'$ is set as $x=b'$
and $A'E'$ is defined as $x'=a'+g(y')$. Instead, in the
transformation way from $AE$ corresponding to $x=a-g(y)$ into $BC$
with $x=b$, the separation is $g(y)+b-a$. Here $(x,y)$ is an
arbitrary point in the virtual space while $(x',y')$ in the
physical space. Practically constants $a'-b'$ or $b-a$ do not
affect the resulting shape of phase front, so the effective OPLs
are identical between the two ways since $y=y'$. Thereupon the ITM
can implement exactly the same function as DTM. In detail, the
slab in Fig.~\ref{Schem}(b) can convert a planar wave front into a
curved one, the opposite shape of $AE$, which is just the function
of $OA'E'D$ in Fig.~\ref{Schem}(a). Meanwhile, within the context
of DTM the slab should be capable of converting a curved wave $AE$
into a plane wave $BC$. Therefore the slab by ITM has a twofold
function which we denote by \emph{I} and \emph{II}, respectively.
For simplicity, we do not consider the transformation in the $y$
direction.

By ITM the spatial distortion can be recorded as a mapping between
the original and transformed spaces,
\begin{eqnarray}\label{transform}
x'=\rho(x,y)\frac{bx}{\Delta}, ~~~y'=y,~~~z'=z,
\end{eqnarray}
where $\rho(x,y)$ is a scaling factor and
\begin{eqnarray}\label{pc}
\Delta=a-g(y)
\end{eqnarray}
corresponding to $AE$, the reversed shape of the wanted wavefront.
The mapping means that the virtual space is stretched or squeezed
proportionally into the physical space. Theoretically one can
choose a transform freely \cite{Pendry2006,Leonhardt2006}. The
present choice is made as $\rho(x,y)=1$ in order that the
parameters change linearly and the slab can be implemented easily.
The difference from the literature is that it has such a
simplicity that it needs not consider the details of
transformation. According to transformation optics, the
permittivity tensor {\boldmath$\varepsilon$} and permeability
tensor {\boldmath$\mu$} in the transformed coordinate system are
connected with the original {\boldmath$\varepsilon_o$} and
{\boldmath$\mu_o$} by the relationships:
$\boldsymbol{\varepsilon}=J\boldsymbol{\varepsilon_o}J^T/
\hbox{det}(J)$,
$\boldsymbol{\mu}=J\boldsymbol{\mu_o}J^T/\hbox{det}(J)$ where
$J^{i'}_{i}={\partial x^{i'}}/{\partial x^{i}}$ is the Jacobian
transformation matrix \cite{Schuring2006a}. Applying
Eq.~(\ref{transform}) gives a general result of the relative
material parameters for the conversion between plane wave and
arbitrarily curved wavefront:
\begin{equation}\label{eu}
\boldsymbol{\varepsilon}=\boldsymbol{\mu}=
\begin{bmatrix} \frac{b}{\Delta}+
\frac{{x}^{2}{{\Delta'}^{2}}}{b\Delta} & -\frac{x\Delta'}{b} & 0\\
-\frac{x\Delta'}{b} & \frac{\Delta}{b} &0\\
0 & 0 &\frac{\Delta}{b}
\end{bmatrix},
\end{equation}
where $\Delta'=\partial{\Delta}/\partial{y}$ and the original
space is considered as vacuum. Notice that the superscripts of
coordinate variables have been omitted for simplicity.  No
singularity exists for {\boldmath$\varepsilon$} and
{\boldmath$\mu$} as long as $AE$ does not intersect with $y$-axis.
Thereupon, the problem of singularity \cite{Tichit2011,Kong2007}
arising mathematically can be avoided completely by taking
appropriate transformation ways. It is important to note that the
complex amplitude transmittance \cite{Goodman1996} of
electromagnetic wave through the slab can be written as
\begin{eqnarray}\label{t}
t=e^{-ik_0\Delta},
\end{eqnarray}
where $k_0$ is the free-space wavenumber and the constant phase
factor has been omitted.

\subsection{The general transformation method for
the transformation between two arbitrary
wavefronts}\label{cformulation}

\begin{figure}\center
\includegraphics[width=4cm]{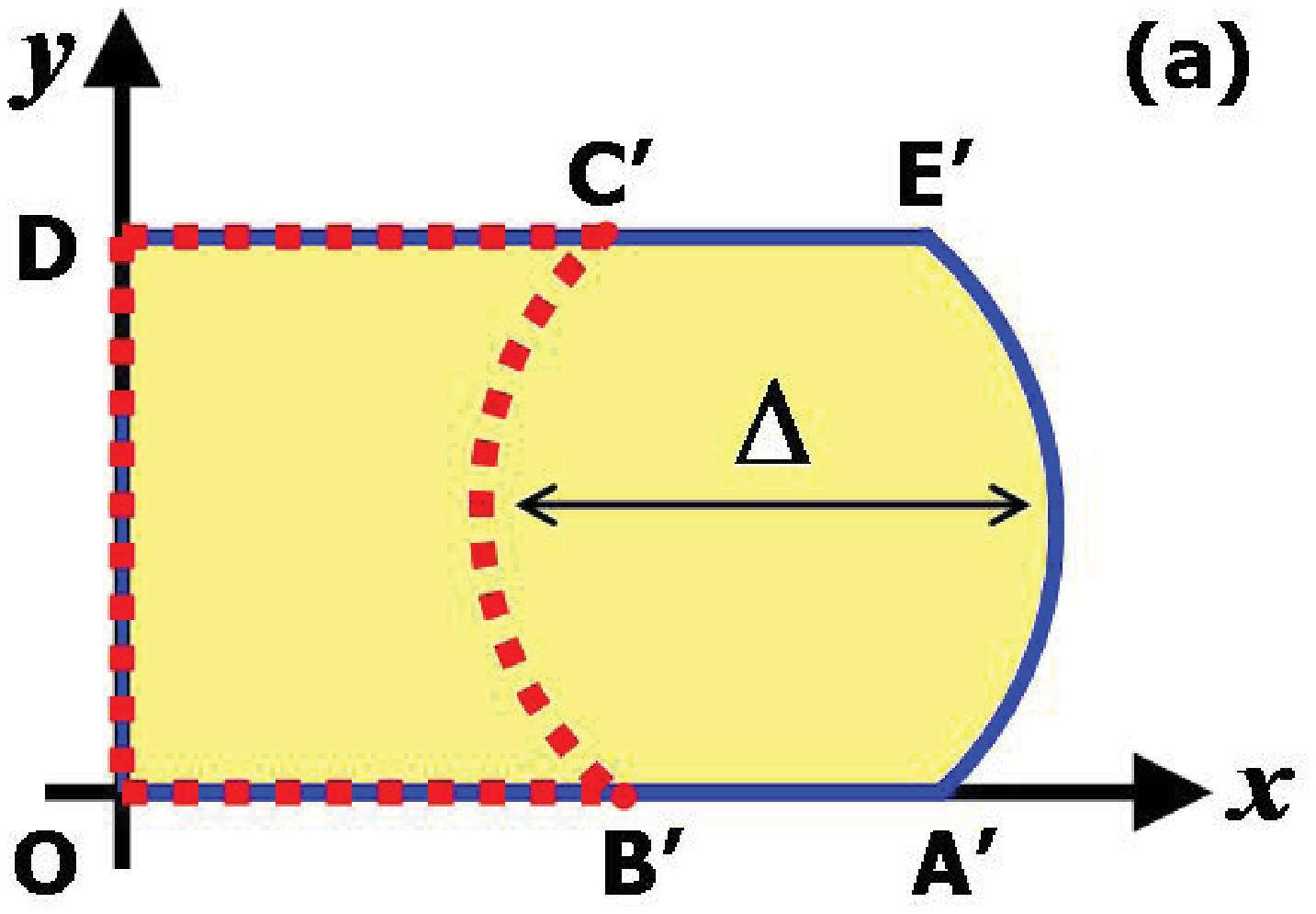}
\includegraphics[width=4cm]{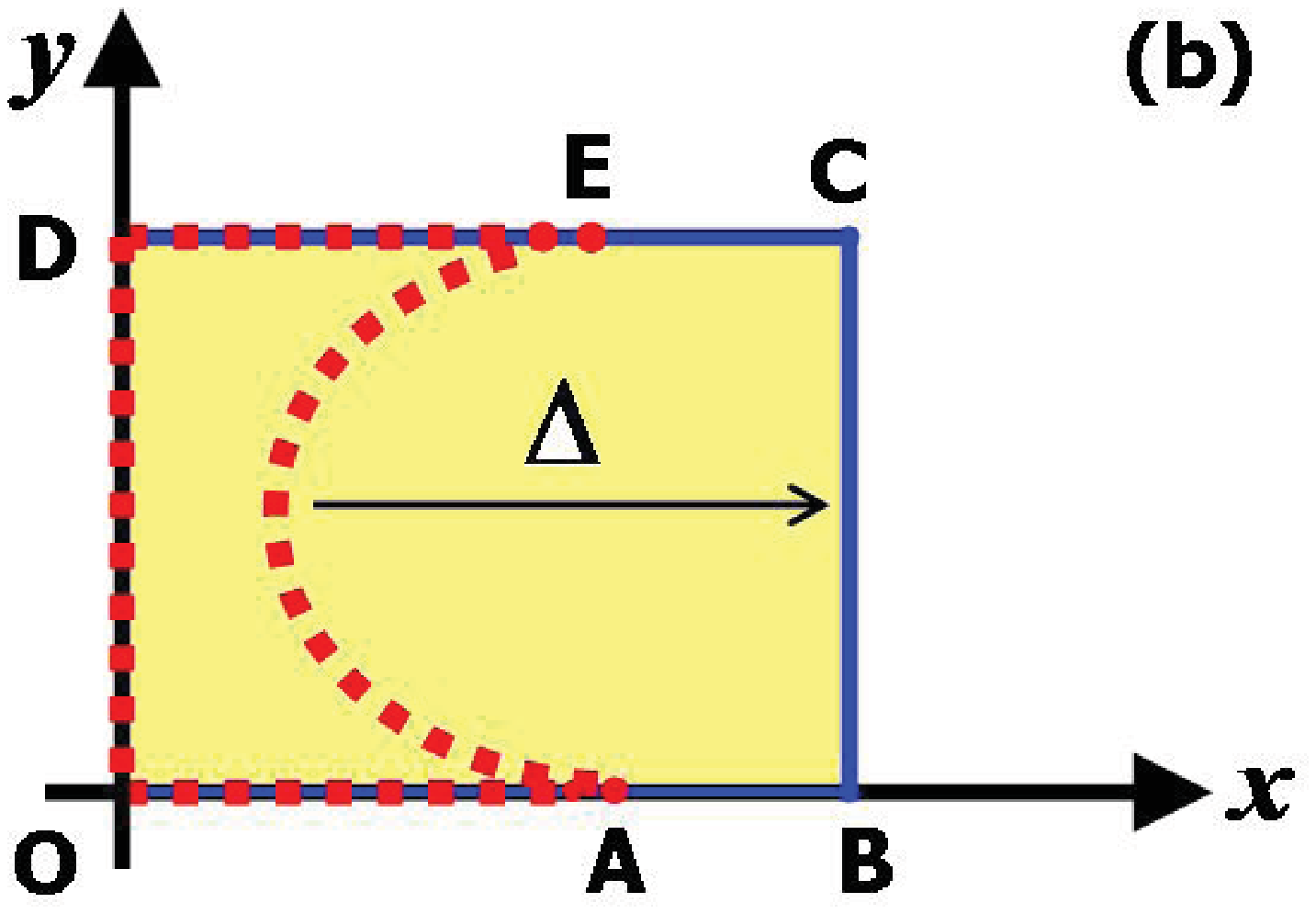}
\caption{\label{Schem2} Schematic diagram of the space
transformation to convert a curved phase front into another curved
one by (a) direct transformation method in literature and (b) the
transformation method in the present work. $AE$, a profile of the
spatial separation of the two wavefronts $B'C'$ and $A'E'$, is
taken to be transformed into a plane.}
\end{figure}

In an analogous fashion to the above discussion, we now examine
how to convert a curved wave front into any desired one by a slab.
Fig.~\ref{Schem2} applies as a sketch wherein $B'C'$ is curved
now. The DTM in literature is to transform a curved wave to the
desired wave shape immediately, i.e. from $B'C'$ to $A'E'$, and
results in designs with irregular shapes, $OA'E'D$. Here, we
generalize the ITM following the same principle based on OPL.
Briefly, one also can imagine compressing $A'E'$ into a plane
$BC$, then the original $B'C'$ would undergo the same distortion
in the $x$ direction becoming $AE$. However, the transformation
becomes that \emph{the profile of the spatial separation between
the original and desired wavefronts is converted to a plane
surface}, i.e. from $AE$ to $BC$, to realize the design by a slab
as $OBCD$ in Fig.~\ref{Schem2}(b).

In transforming any point from $B'C'$ to $A'E'$, the distance
experienced is $g_{2}(y)-g_{1}(y')+a'-b'$, where $B'C'$ and $A'E'$
are supposed to be $x'=b'+g_{1}(y')$ and $x=a'+g_{2}(y)$,
respectively. Instead, in the transformation way from $AE$ with
$x=a+g_{1}(y)-g_{2}(y)$ to $BC$ with $x=b$, the separation
traversed is $g_{1}(y)-g_{2}(y)+b-a$. Actually $a'-b'$ or $b-a$ do
not affect the resulting shape of front, so the effective OPLs are
identical for the two ways since $y=y'$. Therefore the results
Eqs.~(\ref{transform}) and (\ref{eu}) still hold true except that
$\Delta$ now becomes
\begin{eqnarray}\label{cc}
\Delta=a+g_1(y)-g_2(y)
\end{eqnarray}
corresponding to the profile of spatial separation between the two
wavefronts. If the incident wavefront is planar, e.g.
$a+g_1(y)=b'$, then $\Delta=b'-g_2(y)$, which is the reversed
shape of the outgoing wavefront. That is to say, to transform a
plane wave into a curved wavefront by a slab, it only needs to
transform the reversed shape of the latter into the former. This
is just the above conclusion of Eq.~(\ref{pc}), so Eq.~(\ref{cc})
is a general result.

It is important to note that in deriving Eq.~(\ref{pc}) or
Eq.~(\ref{cc}), $AE$ should be shifted so as not to intersect the
$y$ axis to avoid the singularity in $\boldsymbol{\varepsilon}$
and $\boldsymbol{\mu}$ in Eq.~(\ref{eu}). That is, $\Delta$ can be
added with a constant that will not change the generated
wavefront, but only affect the values of
$\boldsymbol{\varepsilon}$ and $\boldsymbol{\mu}$ in the slab. We
will address this point later.

\section{Application of the transformation method}
\label{application}

Applying the above method, a wavefront can be converted into any
desired one through a slab with $\boldsymbol{\varepsilon}$ and
$\boldsymbol{\mu}$ determined by Eq.~(\ref{eu}). In the following
we discuss three kinds of phase transformation to illustrate the
method. Incidentally, similar wavefronts could have been generated
by other media using different methods somewhere in literature,
e.g. conventional lens or transformation media, but our principal
emphasis is on the new general method used to design \emph{flat}
media to induce phase shift and bend wavefront.

\subsection{{Planar to planar wavefront
transformation.\label{secpp}}}

Suppose a plane wave to be deflected into other directions by the
slab. Without loss of generality, the reversed wavefront $AE$ is
chosen as $y=-(x-a)\cot{\theta}$ such that the exit wave is
deflected clockwise  by an amount $\theta$. Then
\begin{eqnarray}\label{deflectwavefront}
\Delta=a-y\tan{\theta}
\end{eqnarray}
and using Eq.~(\ref{eu}) one obtains the relative material
parameters for the slab.

\begin{figure}\centering
\includegraphics[width=4cm]{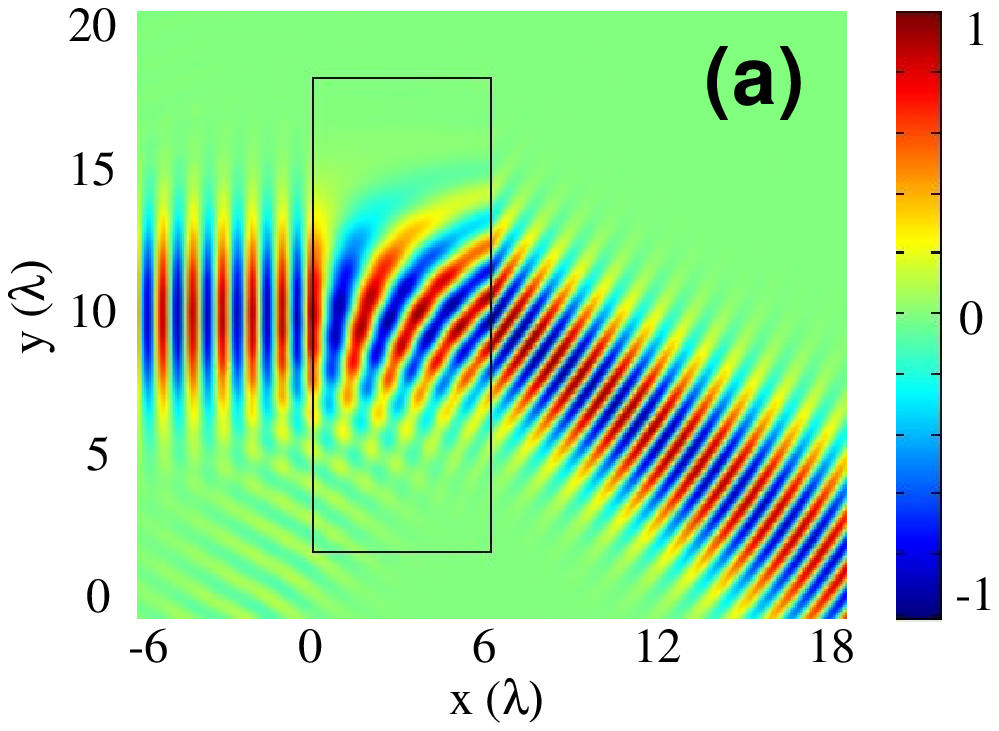}
\includegraphics[width=4cm]{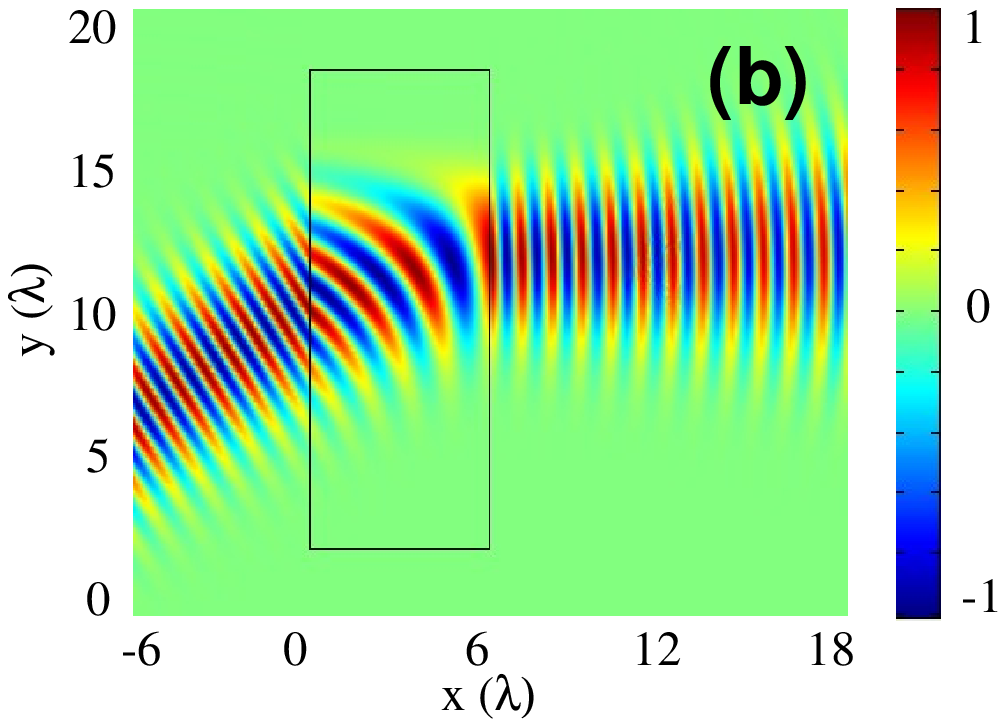}
\includegraphics[width=4cm]{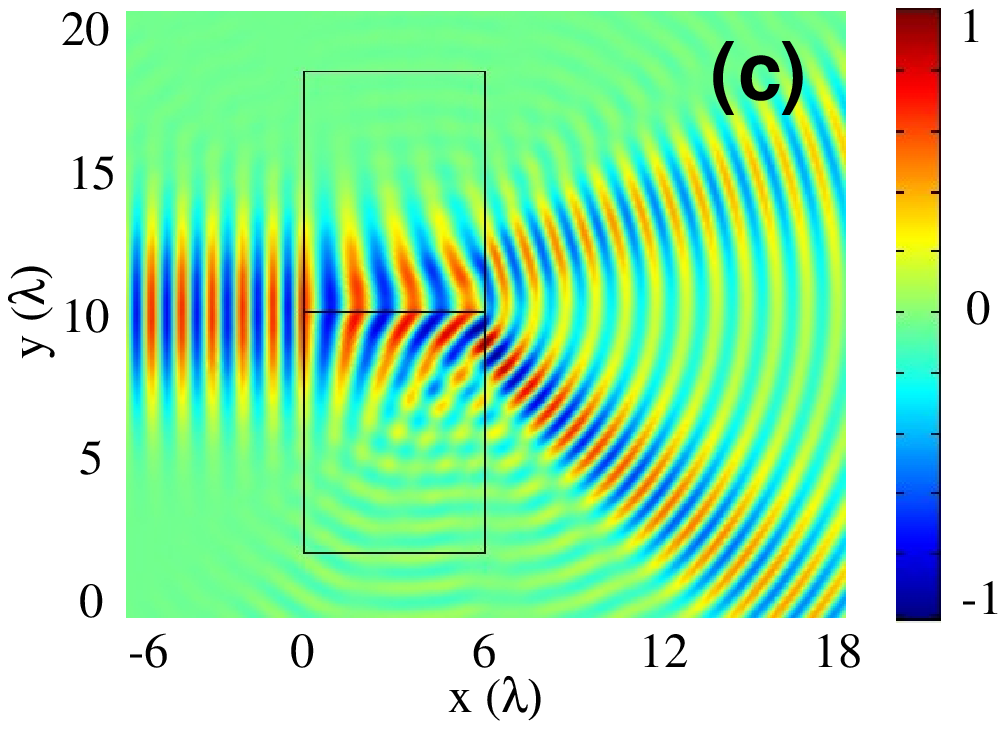}
\includegraphics[width=4cm]{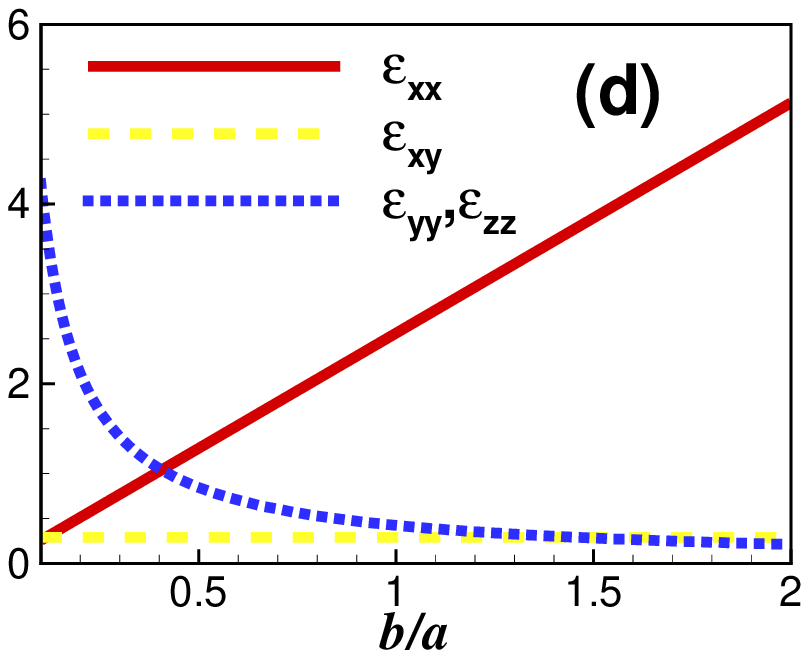}
\caption{\label{planar} Normalized $E$-field patterns in the slab
deflector for (a) normal incidence with exit angle $\pi/6$ and (b)
incident angle $-\pi/6$ with normal outgoing, and (c) in a
splitter formed by the upper slab with deflection angle of
$-\pi/12$ and the lower of $\pi/6$. (d) The permittivity at the
center of the slab varies with the width $b$.}
\end{figure}

To test the effectiveness of the ITM and  the  proposed design,
suppose $a=0.4$m, $b=0.3$m and $\theta=\pi/6$. Consider a
$z$-polarized Gaussian beam with a wavelength of $\lambda=0.05$m
incident on the slab of $d=0.8$m (the choices are uniform
throughout), the electric field distribution obtained from a
finite-element simulation is shown in Figure~\ref{planar}.
We see that the incident wavefront is gradually tilted inside the
slab until the deflection result is transferred out to the free
space. The exit beam acquires a new direction of propagation
unlike that of the incident one and the deflected wavefront is not
necessarily parallel to the exit interface. Interestingly, by
comparing (a) and (b) we find that the beam is deflected by the
same amount whether for normal or oblique incidence. The
underlying mechanism is that this transformation is in fact an
operation of rotation on the local coordinate system which will be
imposed on any fixed incident field vectors, resulting in wave
vectors with identical deflection. This fact confirms the twofold
function of the slab: \emph{I}-Convert a normally incident plane
wave into another with the deflected angle $\theta$;
\emph{II}-Convert an oblique plane wave with the incident angle
$-\theta$ into a normal one. Also the ITM is justified.

To change the deflection amount, one can adjust the parameter
$\theta$. Positive or negative $\theta$, i.e. clockwise or
counter-clockwise deflection, even extreme angles approaching the
boundary, can be realized. Alternatively, one can realize
different deflections by stacking certain pieces of slabs with
given deflection amounts. In addition, combine two such slabs and
a beam splitter is created, e.g. Fig.~\ref{planar}(c). A benefit
of this splitter is that a beam can be split into any two
directions, provided that the deflection angles are set
accordingly. We also find that the width $b$ can be chosen freely,
so the slab can be made as compact as desired only at the expense
of a larger range of material parameters, as shown in
Fig.~\ref{planar}(d).

\subsection{Planar to curved
wavefront transformation.}\label{secpc}

For example, we consider how to convert a plane wave to a
diverging/converging wave.  $AE$ is assumed to be a paraboloid
\begin{eqnarray}\label{Gaussianwavefront}
\Delta=a-{y^2}/{2R(a)}
\end{eqnarray}
positioned at $x=a$ where the radius of curvature
\begin{eqnarray}\label{Gaussianradius}
R(x)=\pm x(1+{x_0}^2/{x}^2)
\end{eqnarray}
and the Rayleigh range $x_0=w_0^2{\pi}/{\lambda}$ corresponding to
a Gaussian beam of waist radius $w_0$.
Eqs.~(\ref{Gaussianwavefront}) and (\ref{Gaussianradius}) will be
of continuous use from now on. Applying the ITM, the inverse
transformation is taken from a converging/diverging surface to a
planar one. Then the material parameters are obtained by using
Eq.~(\ref{eu}). Similar transformations were used to generate
plane waves \cite{Kwon2008a,Lin2008a}, but here we make use of the
map inversely.

\begin{figure}\centering
\includegraphics[width=4cm]{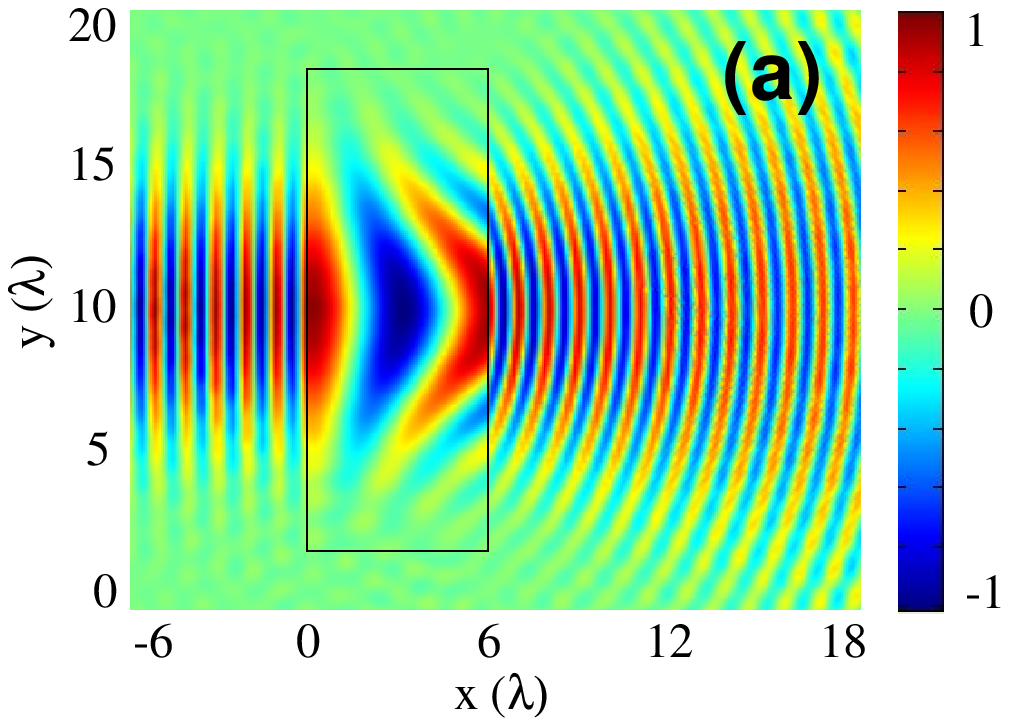}
\includegraphics[width=4cm]{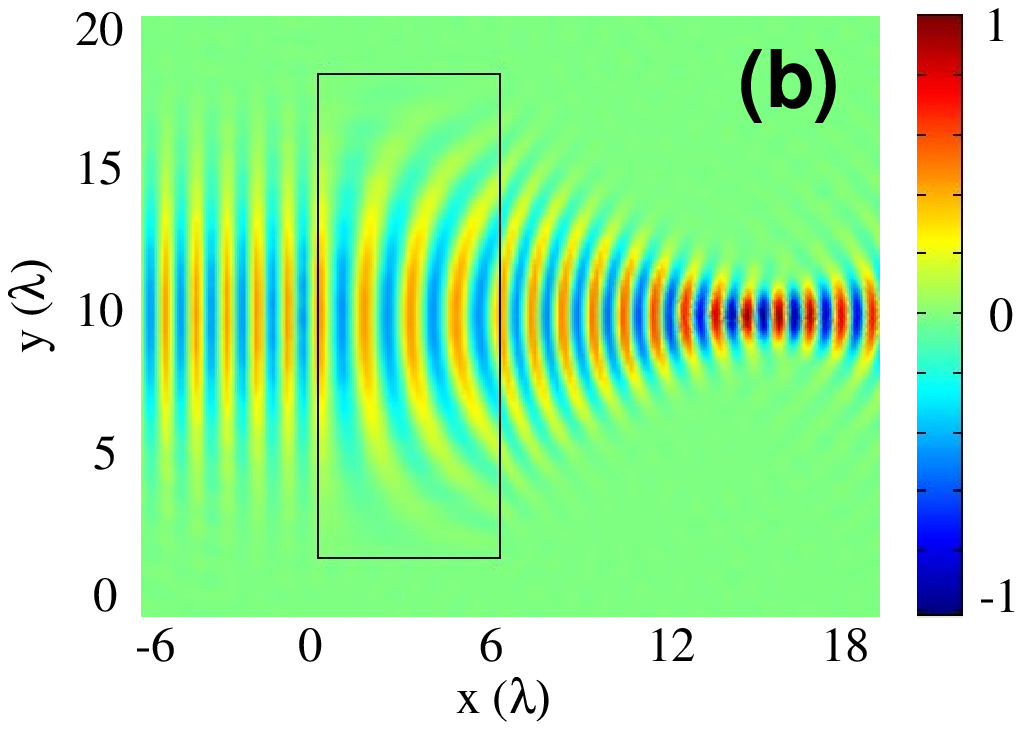}
\includegraphics[width=4cm]{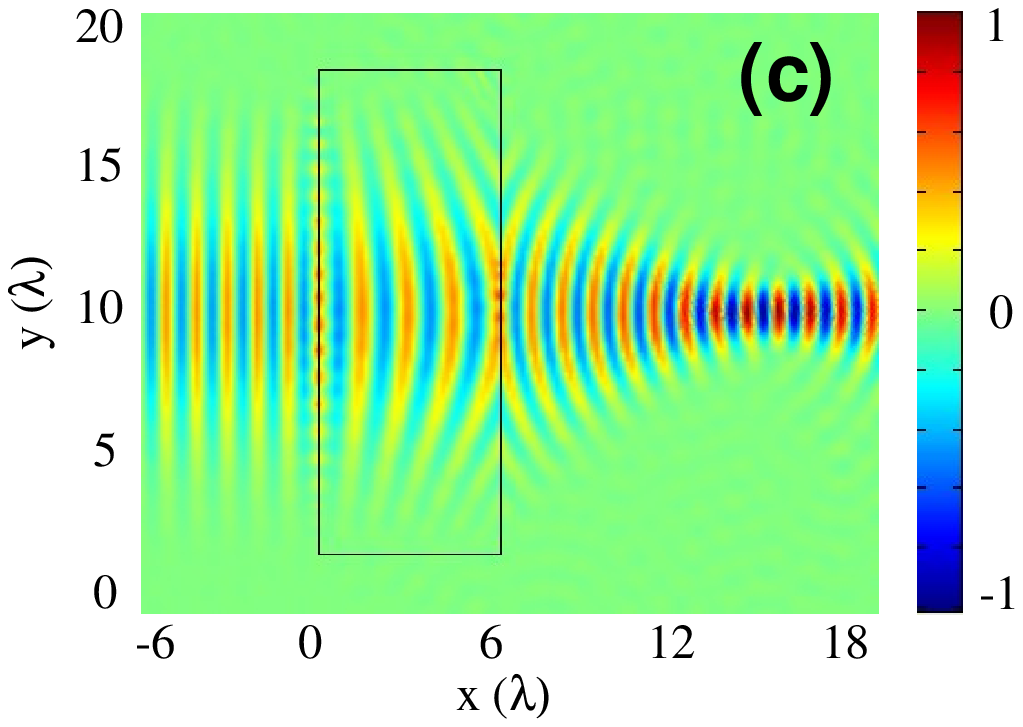}
\includegraphics[width=4cm]{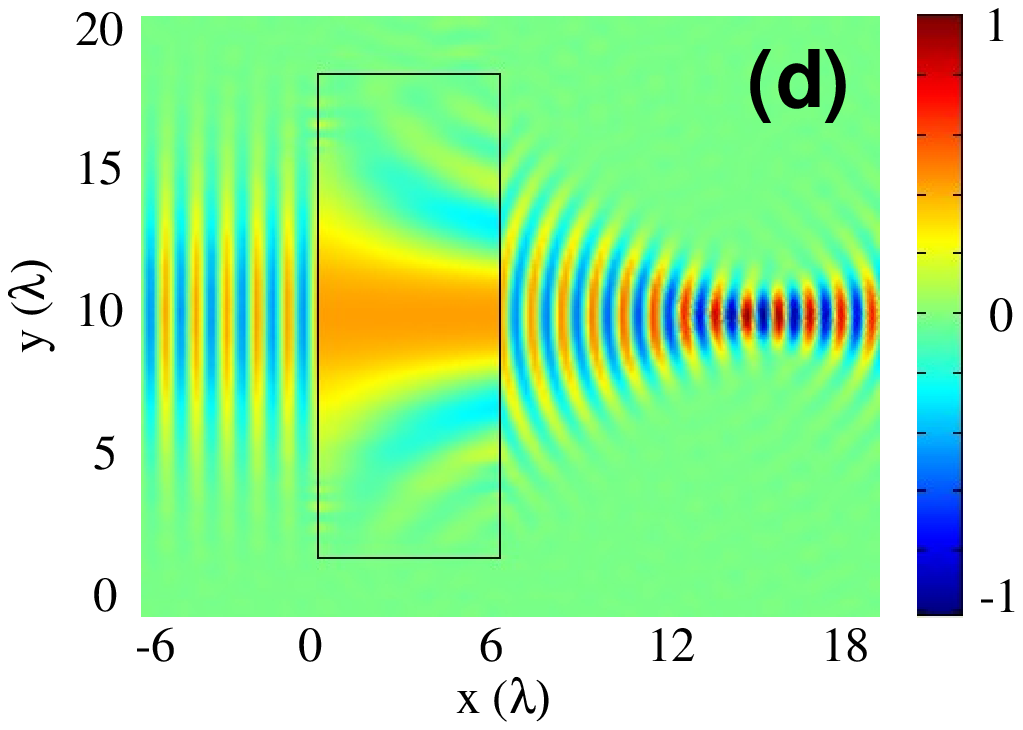}
\caption{\label{pcurve} Normalized $E$-field patterns in the slab
applying (a) \emph{converging}/(b,c,d) \emph{diverging} surface to
planar one transformation. The slab can convert a planar wavefront
to a (a) \emph{diverging}/(b,c,d) \emph{converging} one. The
surfaces to transform are chosen as $\Delta=0.05+(y-0.5)^2$ in
(a), $0.2-(y-0.5)^2$ in (b), $-0.2-(y-0.5)^2$ in (c), and
$-(y-0.5)^2$ in (d). The incident beam centers are $(0,0.5)$ and
the waists $w_0$ are $0.2$m.}
\end{figure}

In order to show the phase evolution clearly, we choose
$a=0.05$m/$0.2$m. The other values are $w_0=0.2$m,
$R=-0.5$m/$0.5$m and the waist center $(0,0.5)$ for the incident
beam which approximates a planar wavefront, and the simulated
results are shown in Fig.~\ref{pcurve} (a)/(b).
Note that we have adopted the convention that a converging
wavefront has a negative radius of curvature whereas a diverging
wavefront has a positive radius of curvature. In the slab one can
see the incident planar wavefront gradually evolves into a
diverging/converging one at $x=b$, i.e., the function \emph{I}.
The central area of the wavefront travels faster/more slowly than
its outer edges and then the midregion overtakes/lags behind those
edges, thereby continuously bending the wavefront. In terms of
Fermat's principle, it means that the refractive index is
lower/larger in the central than in outer edges. In
Fig.~\ref{Schem} the generated surface $A'E'$ is the mirror image
of the transformed surface $AE$, so one can see that the radius of
exit phase surface is $0.5$m/$-0.5$m just contrary to $R$ and the
new waist varies accordingly.

On the other hand, the slab in (a)/(b) enables \emph{II}-make a
converging/diverging wavefront bend into a planar one, as shown in
the literature applying similar coordinate transformations
\cite{Zhang2008,Kwon2008a,Lin2008a,Tichit2011}. Hence the slab
also has a twofold function which may find import applications.
For instance, it is capable of implementing flat lens. Recall that
for a thin lens the complex amplitude transmittance is
$t=e^{ik_0nb_0-ik_0y^2/2f}$, where  $n$ is the refractive index,
$b_0$ is the thickness of the center, and $f$ is the focal length
\cite{Goodman1996}. Therefore, the slab determined by
Eq.~(\ref{Gaussianwavefront}) is equivalent to a thin lens with
the focal length $f=R$, which was confirmed by Fig.~\ref{pcurve}.
Apart from its shape being a compact slab, such a lens has another
advantage that the location of focus can be tuned continuously by
changing the material parameters or discretely by stacking certain
pieces of slab with unit focus distances. Note that the result in
(b) is very similar to but nevertheless distinct from previous
results of flattening lens \cite{Roberts2009,Yang2011}. The
transformation in the latter was to reshape a known optical
component, whereas the present method immediately take the
wavefront to transform and thus is endowed with great flexibility,
so results $\boldsymbol{\varepsilon}$ and $\boldsymbol{\mu}$ are
different, too.

To examine the role the factor $a$ plays in the result, we alter
its value in (b) as $a=-0.2$m and $0$m, and the results are shown
in (c) and (d), respectively. Note that the original wavefront in
(c) and (d) are chosen left to the slab, whereas that in (b) is
inside the slab. Comparing (b) and (c) or (d), one can readily see
that the period number of wave inside the slab equals $a/\lambda$.
This fact can be understood by Fig.~\ref{Schem} where the original
wave distribution in $OAED$ is transformed into that in $OBCD$ and
$OA=a$, so the period number of the original wave along $OA$ is
just that in the slab along $OB$.  Particularly interesting is the
crest in (d) that becomes quite elongated and stretches into a
long beam inside the slab, exhibiting a extremely large
wavelength. Correspondingly, the index of refraction should be
nearly zero \cite{Ziolkowski2004}, which will be proved later.

At the same time, it can be seen from (b), (c) and (d) that the
exit wavefronts remain fixed, converging at a point $R$ away from
the slab. This fact results from that the curvature of radius in
the transform is uniform. So they can implement the same function
and thus the original surface $AE$ in Fig.~\ref{Schem} need not be
chosen in the slab.  Particularly, in contrast with (b), it seems
like that the central region of wavefront travelled faster than
the outer edges inside the slab in (c) and (d), like in (a).
However, the fact is quite the contrary and the true reason is
that negative refraction occurs in (c) and (d), thereby
introducing negative phase shift. Since the central area of the
beam travels more slowly than its outer edges, the subtracted
phase is less and the accumulated phase of the midregion is
advanced with respect to those edges on the back surface, whereby
the outgoing wavefront are deformed as in (b). This point is
evident from the opposite phases traversing the boundary in (c)
and (d). Such a process leads to the \emph{reversal of phase} that
may find important applications, such as wavefront correction in
imaging or adaptive optics. In contrast the phases across the
interfaces are uniform in (a)/(b) and the refractive indexes are
positive. Therefore, we come to an important conclusion that the
ways to get a desired wave are diverse and then the same to the
choices of slab, thereby increasing the possibility of
realization.

\subsection{{Curved to curved wavefront transformation.\label{seccc}}}

\begin{figure}\centering
\includegraphics[width=4cm]{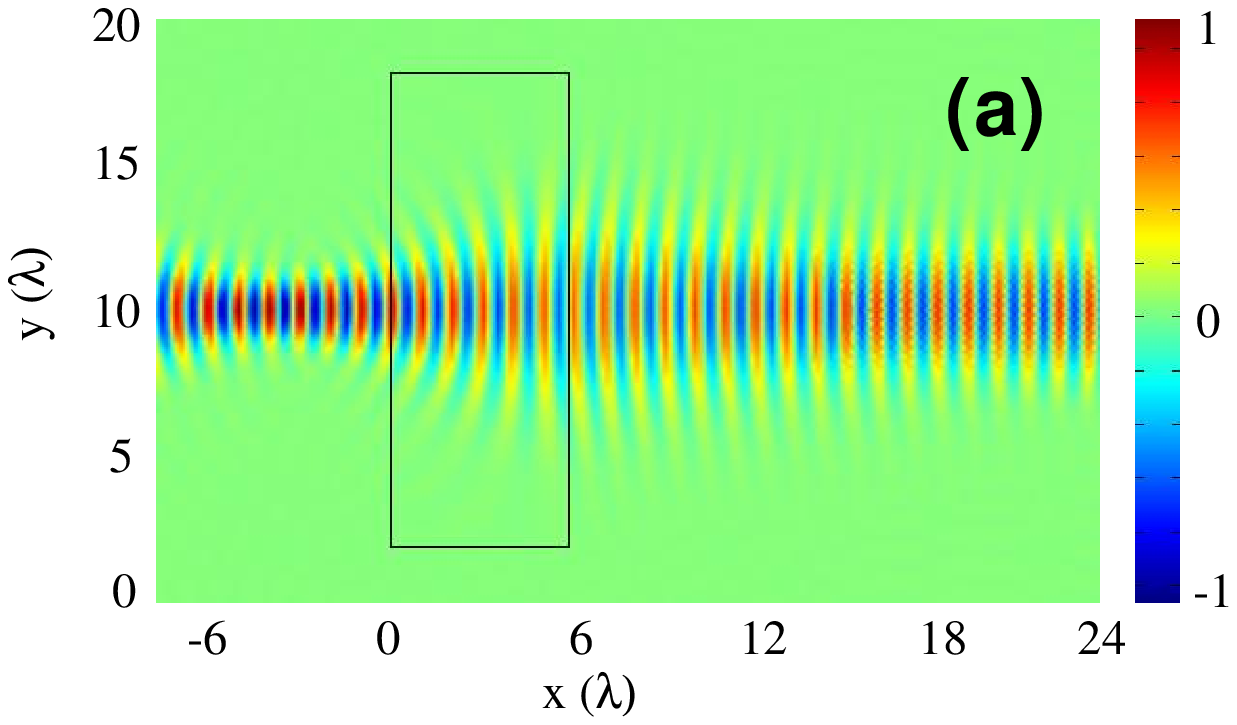}
\includegraphics[width=4cm]{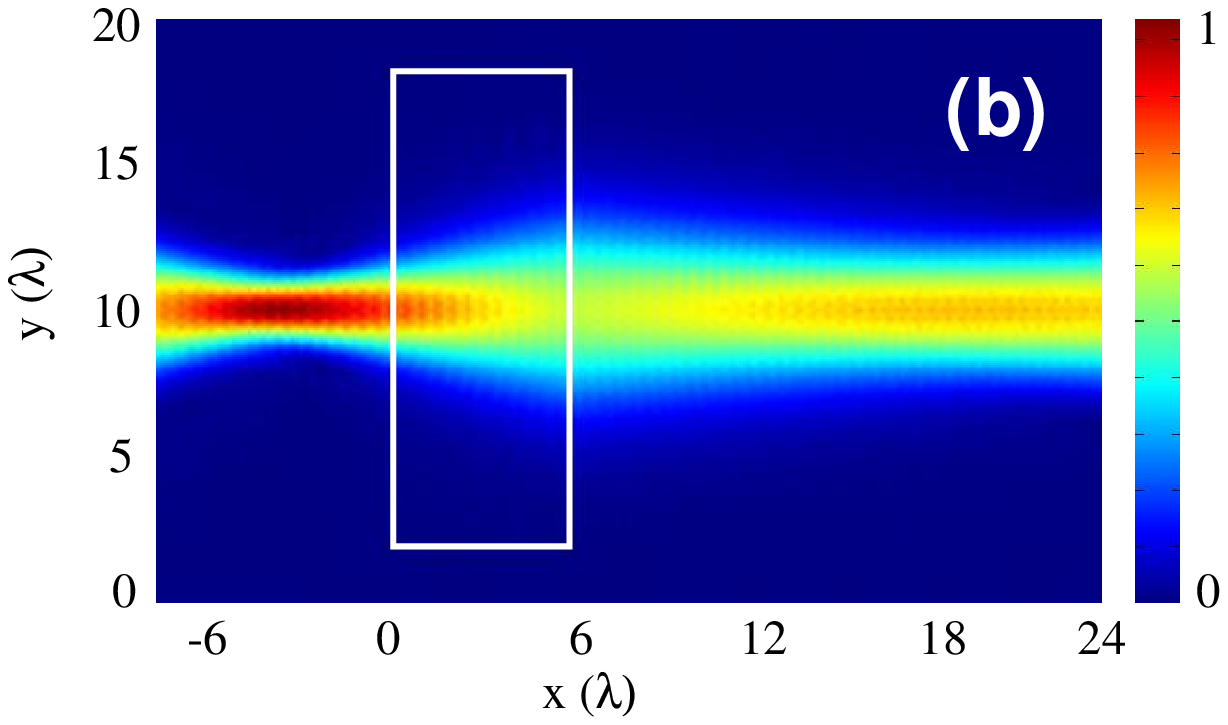}
\includegraphics[width=4cm]{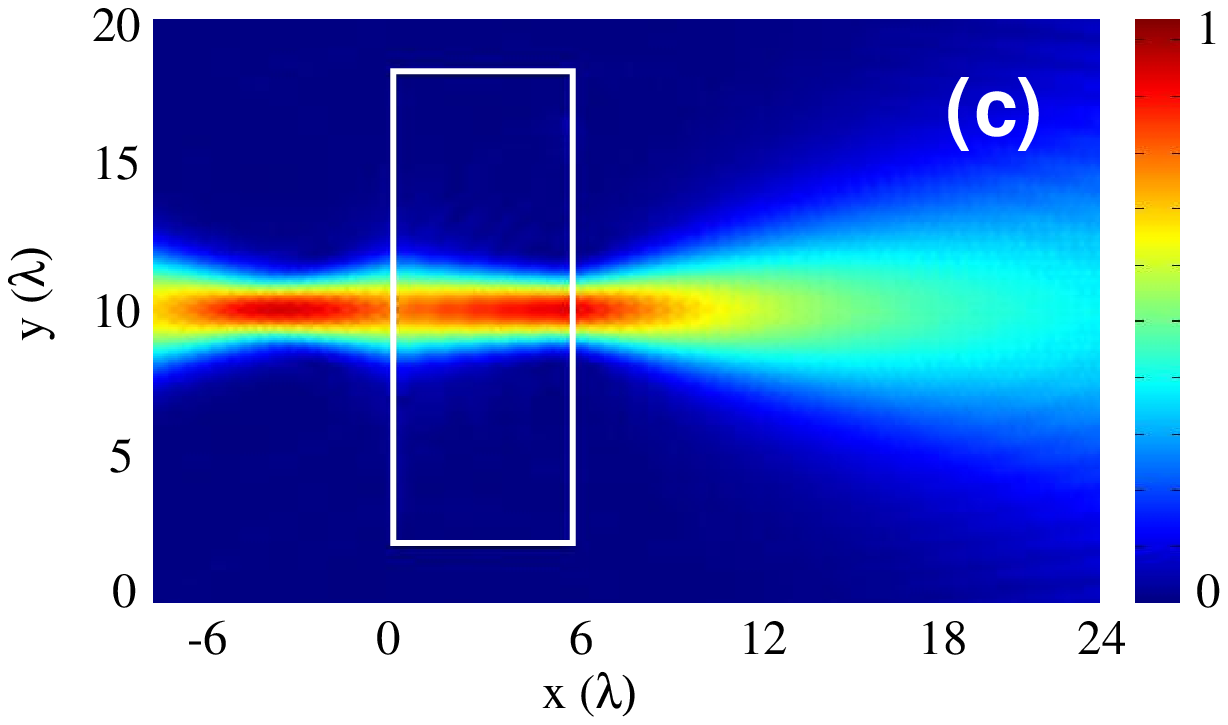}
\includegraphics[width=4cm]{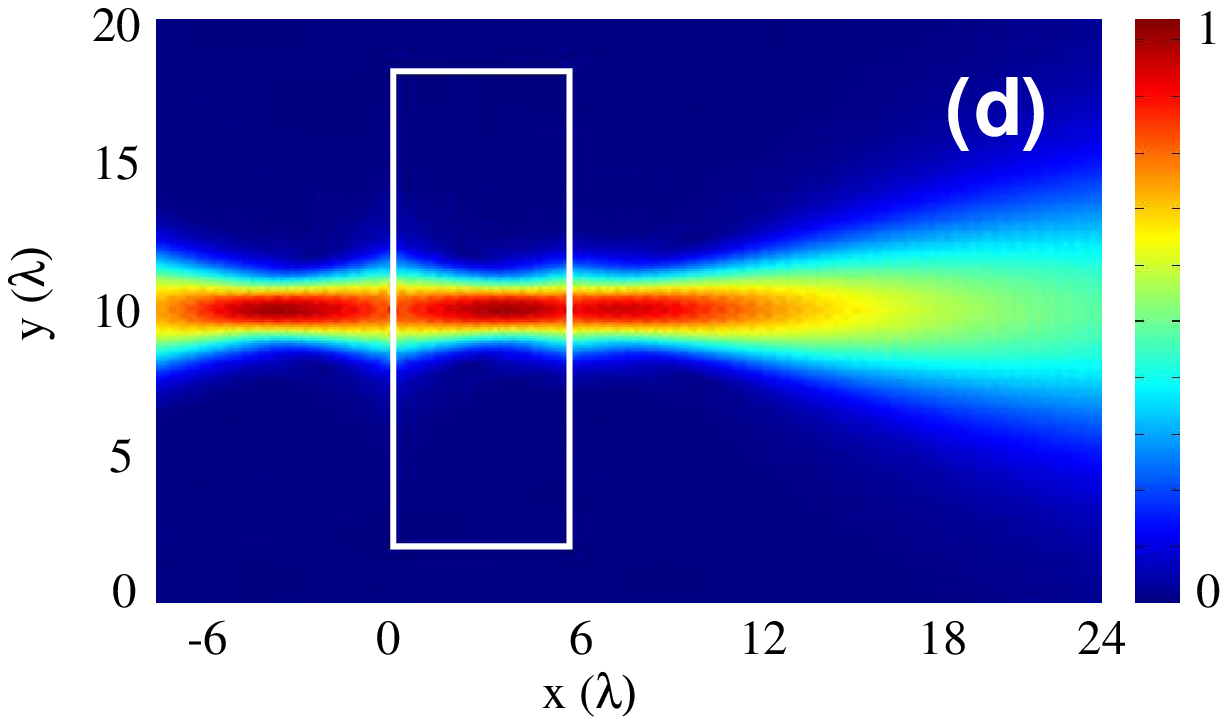}
\caption{\label{ccurve} Normalized  $E$-field patterns (a) and the
energy density (b,c,d) in the slab transforming a diverging
wavefront to a converging one. (c) displays the focus on the exit
boundary when $-a$ equals to the distance of beam center from the
slab. If $-a$ becomes larger, then another focus will appear
inside the slab, leading to focusing of light in (d). The two
wavefronts are chosen as (a,b) $a+g_1(y)=0.3-(y-0.5)^2$, (c)
$a+g_1(y)=-0.15+(y-0.5)^2$, (d) $a+g_1(y)=-0.3+(y-0.5)^2$, while
$g_2(y)=(y-0.5)^2/2.4$ and incident beam center $(-0.15,0.5)$.}
\end{figure}

To be specific we investigate the conversion of a diverging wave
$x=a+g_1(y)$ with $a=0.3$m, $R=0.5$m and the beam center
$(-0.15,0.5)$ to a converging one $g_2(y)$ with $R'=-1.2$m, where
Eqs.~(\ref{Gaussianwavefront}) and (\ref{Gaussianradius}) have
been applied. Then the material parameters are obtained by using
Eqs.~(\ref{eu}) and (\ref{cc}), and the simulated results are
shown in Figs.~\ref{ccurve}(a) and (b). Apparently the left
diverging wave traversing the slab is deformed into a converging
one. As in the above subsection, the period number of wave inside
the slab equals $a/\lambda$ and $R'$ determines the generated
wavefront radius of curvature. At the same time, the slab induces
a transmittance the same as that of a thin lens with the focal
length satisfying $1/f=1/R-1/R'$. Moreover, the focus can be tuned
continuously by changing the parameters of transformation. For
example, let $-a$ equals to the distance of beam center from the
slab, then the beam will focus at the exit surface, as shown in
Fig.~\ref{ccurve}(c). The reason is that the choice of
transformation is to let the wavefront at $-a$, i.e. incident beam
center, be converted to a plane surface. The slab now acts as a
\emph{phase compensator} that reverses the amount of phase change
within the region $a$ away from the slab. Therefore, it is
expectable that, if $-a$ is larger than the distance of beam
center from the slab, then the beam will \textit{focus} inside the
slab, as shown in Fig.~\ref{ccurve}(d). Such a function can be
utilized as beam relaying and flat lens
\cite{Kundtz2010,Roberts2009,Yang2011}.

As known beam expansion or compression has been achieved by
manipulating the longitudinal energy flow
\cite{Xu2008a,Rahm2008b}, that is, expanding or squeezing the
virtual space in the $y$ direction in Fig.~\ref{Schem}. In
contrast, the same task can be performed now by the transformation
on the transverse phase, that is, distorting the virtual space in
the $x$ direction. From the above, the wavefront with the radius
of curvature $R$ at $a$ is transformed into another one with the
radius of curvature $R'$ at $b$. The widths of the two waves, $w$
and $w'$ across the back interface can be considered unchanged.
Then, for the outgoing wave, in accordance with $R'$ and $w'$, the
beam center $x$ away from the slab and the waist radius $w'_0$ can
be found, i.e., $x=-R'/[1+(\lambda R'/\pi w'^2)^2]$ and
$w'_0=w'^2/[1+(\pi w'^2/\lambda R')^2]$. Therefore, the width of
the resultant beam can be engineered through choosing appropriate
wavefront to transform and $R'$. In Figs.~\ref{ccurve}(a) and (b),
the wavefront transformed with a distance from the center
$x=0.45$m has a width $w=w_0\sqrt{1+(x/x_0)^2}=0.15$m, so the
waist radius of the resulted beam equals to $w'_0=0.1$m, two times
the original one. Evidently, the incident beam is expanded into
another one with wider waist. The contrary condition, beam
compression, can be realized similarly.

\subsection{Influence of the coordinate transform on
material parameters.\label{seceu}}

\begin{figure}\centering
\includegraphics[width=4cm]{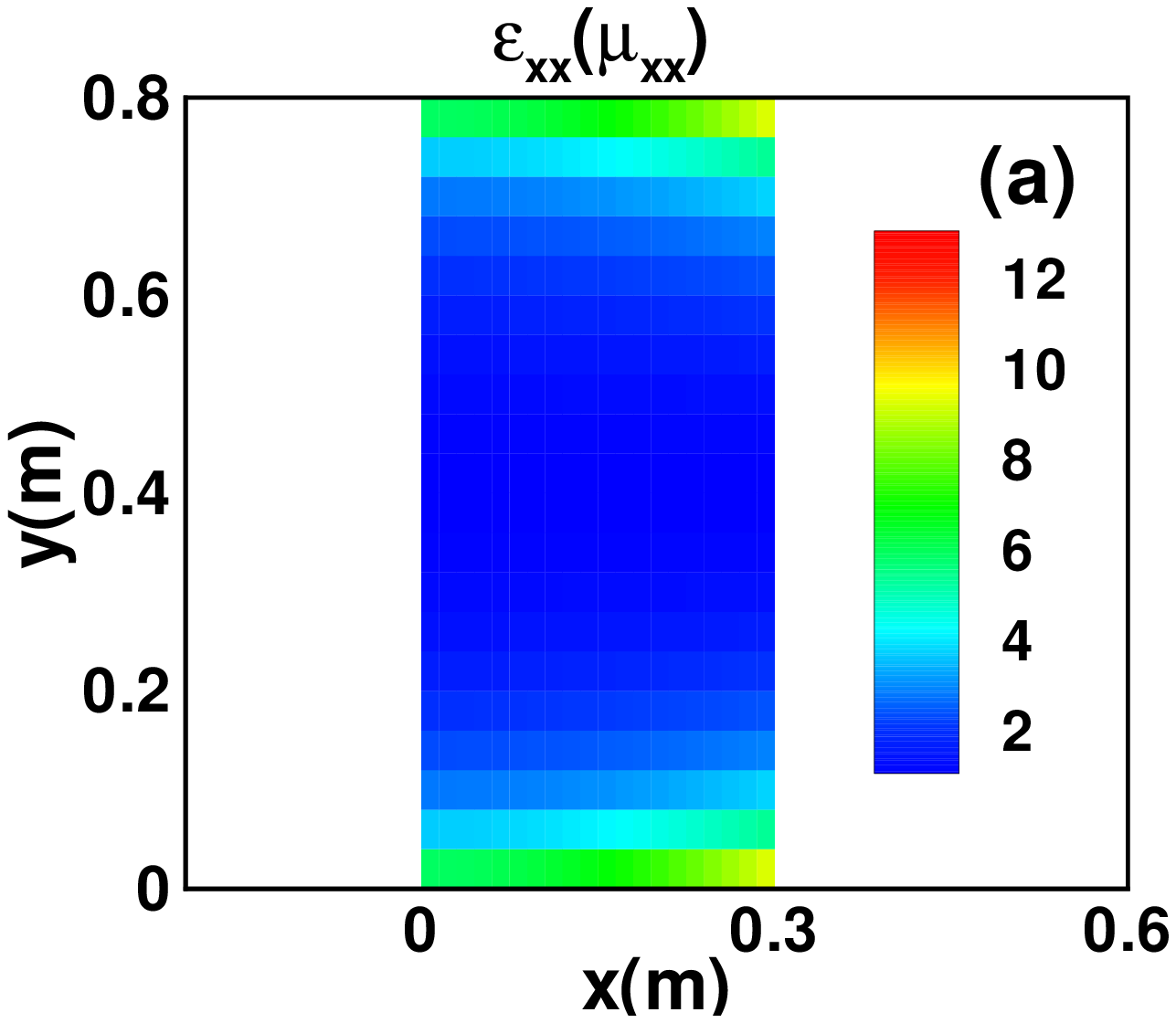}
\includegraphics[width=4cm]{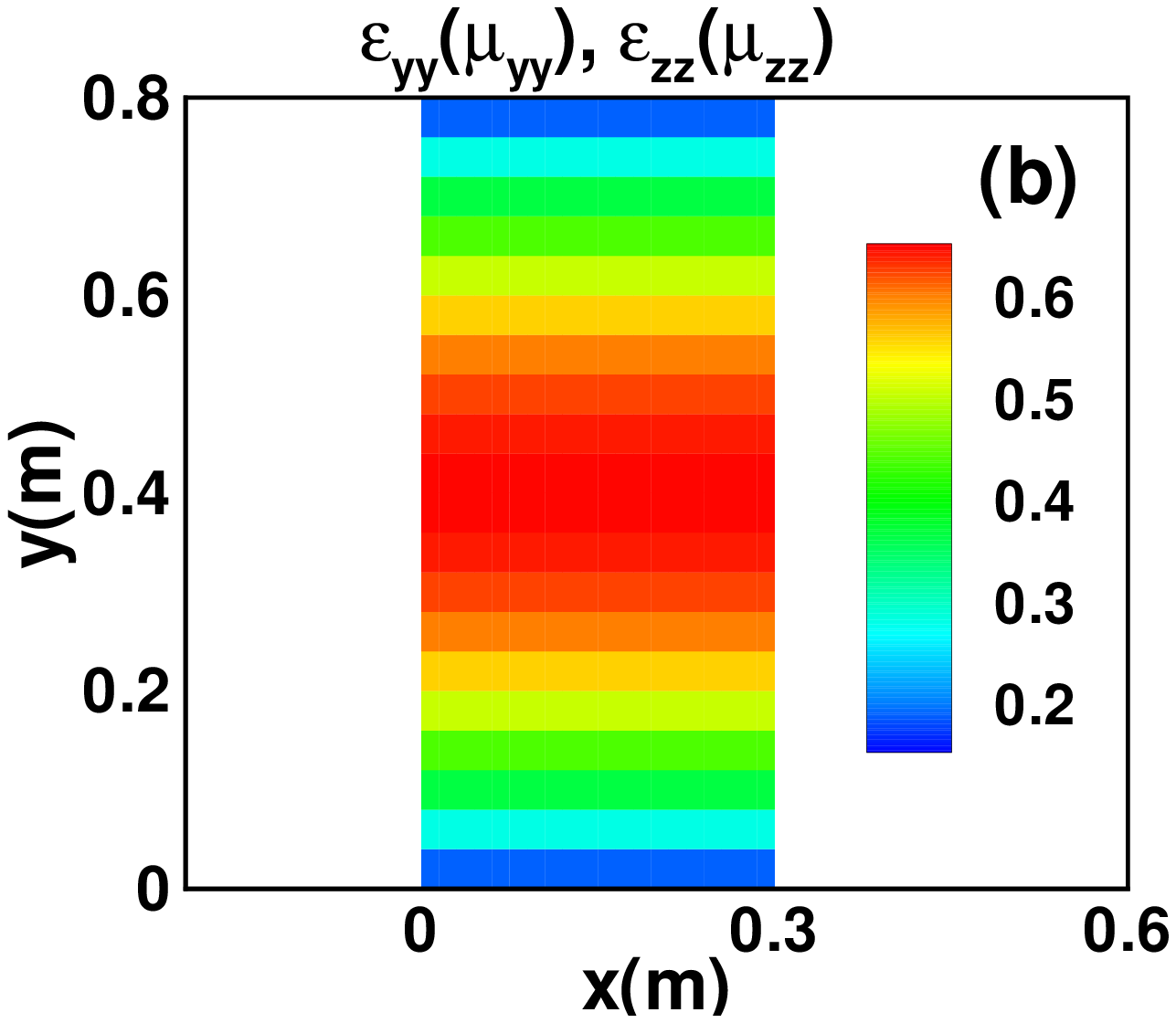}
\includegraphics[width=4cm]{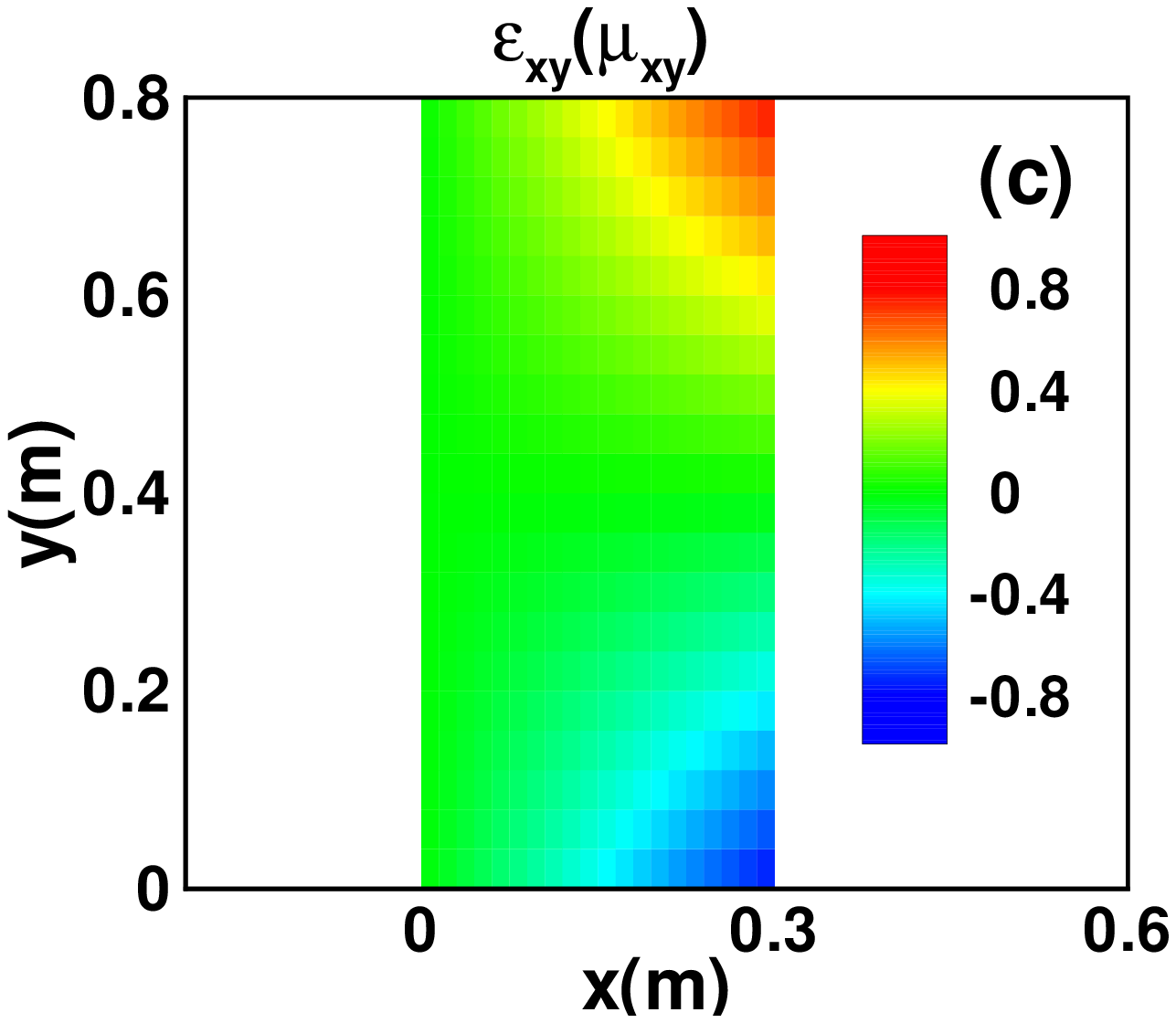}
\includegraphics[width=4cm]{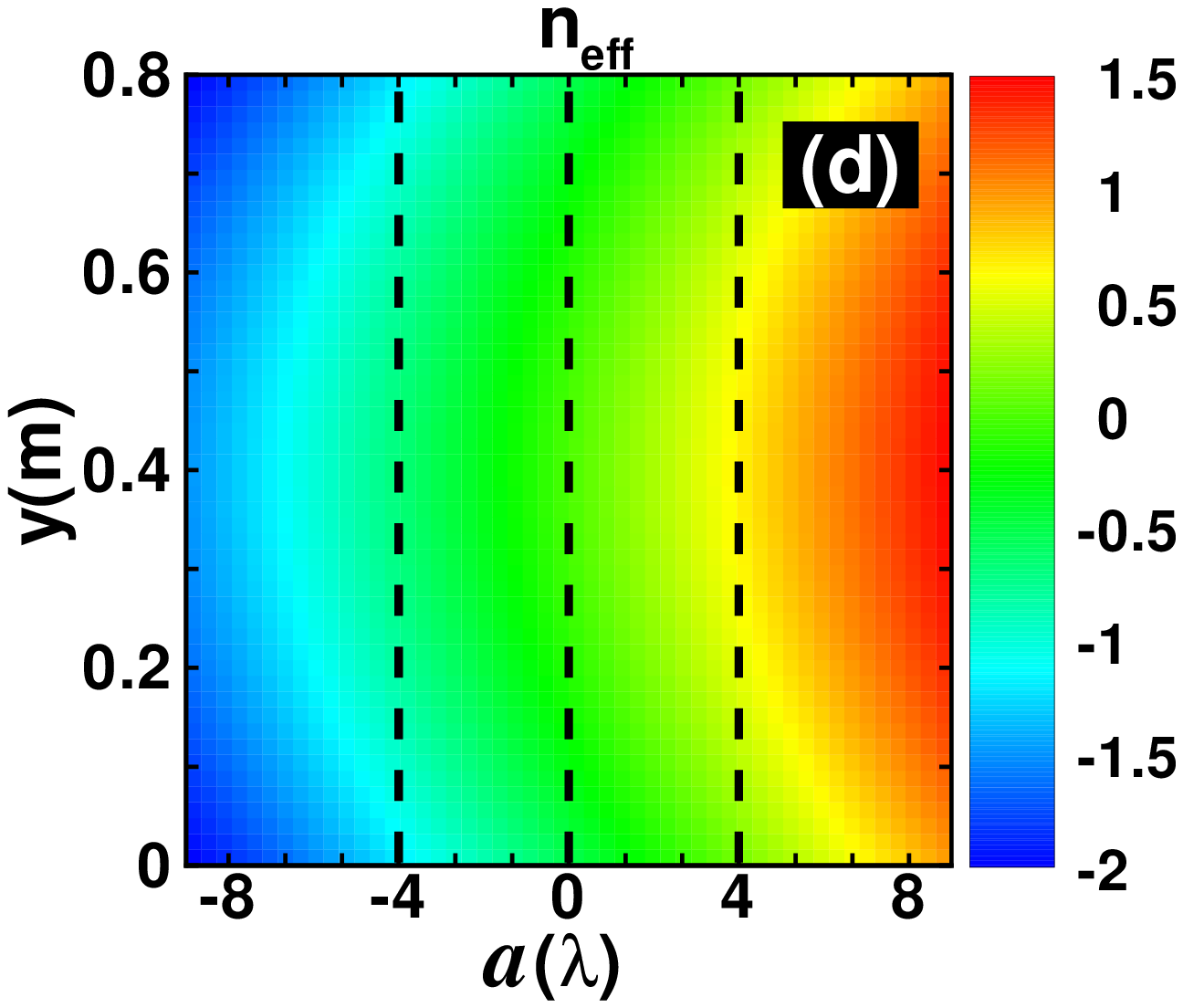}
\caption{\label{parameter} (a), (b) and (c) respectively
corresponds to $\varepsilon_{xx}(\mu_{xx})$,
$\varepsilon_{yy}(\mu_{yy})$/$\varepsilon_{zz}(\mu_{zz})$ and
$\varepsilon_{xy}(\mu_{xy})$. (d) shows the effective refractive
index on the transverse of the slab for different choices of $a$
with $b=0.3$. The three dashed lines in (d) from left to right
correspond to $n_{eff}$ for slabs in Figs.~\ref{pcurve}(c), (d),
and  (b) respectively.}
\end{figure}

Next we discuss the effect of the choice of transformation on the
material parameters. As the above analysis indicates, e.g.
Fig.~\ref{planar}(d), the values of $\boldsymbol{\varepsilon}$ and
$\boldsymbol{\mu}$ change as the slab thickness varies. On the
other hand, the sign and magnitude of $\Delta$ also plays an
important role in determining $\boldsymbol{\varepsilon}$ and
$\boldsymbol{\mu}$. We have known from Sec.~\ref{formulation} that
to avoid the singularity in $\boldsymbol{\varepsilon}$ or
$\boldsymbol{\mu}$, $AE$, i.e. $\Delta$, can be shifted some
amount along the $x$ axis. Despite no impact on the generated
wavefront, it undoubtedly influences the values of
$\boldsymbol{\varepsilon}$ and $\boldsymbol{\mu}$ in the slab.
From Eq.~(\ref{eu}) we conclude that: (i) The larger the magnitude
of $\Delta$, the larger the range of values of
$\boldsymbol{\varepsilon}$ and $\boldsymbol{\mu}$.  In terms of
the principle of the equality of OPL, wave need to travel longer
path $\Delta$ to generate the phase difference and then a
width-fixed slab has to increase the magnitudes of
$\boldsymbol{\varepsilon}$ and $\boldsymbol{\mu}$. (ii) If
$\Delta$ becomes negative, $\boldsymbol{\varepsilon}$ and
$\boldsymbol{\mu}$ will also take negative values. Especially, if
$\Delta$ becomes zero, $\boldsymbol{\varepsilon}$ and
$\boldsymbol{\mu}$ would approach infinity. In this case, $\Delta$
can be added a positive constant. As an example, we present the
constitutional parameters in Figs.~\ref{parameter} (a)-(c)
corresponding to the slab in Figs.~\ref{pcurve}(b).

Further, in order to characterize the slab more clearly and
commonly we introduce the effective average refractive index
\cite{Saleh2007,Goodman1996,Li2008,Tang2010} in terms of the
complex amplitude transmittance Eq.~(\ref{t}),
\begin{equation}\label{n}
n_{eff}=\frac{\Delta}{b}.
\end{equation}
It indicates that the anisotropy is reduced such that nondiagonal
elements in $J$ are eliminated by choosing orthogonal grids in
virtual space to transform. Then the slab can be considered as
nonmagnetic and be realized by dielectrics only though the
performance may be sacrificed a bit \cite{Li2008,Tang2010}. The
effective refractive index on the transverse of the slab for
different choices of $a$ are shown in Fig.~\ref{parameter}(d)
which has included the results in Figs.~\ref{pcurve}(b), (c) and
(d). It shows that $n_{eff}$ increases with $a$ and that $n_{eff}$
in the center region is larger than in the outer area. This is
reasonable because the larger $a$, the more original space will be
squeezed into the slab in Fig.~\ref{Schem}(b). For the midregion,
if $a\leq 0$, then $n_{eff}\leq 0$, e.g., the left two lines in
Fig.~\ref{parameter}(d) corresponding to $n_{eff}$ for slabs in
Figs.~\ref{pcurve}(d) and (c), respectively. Accordingly, the
choice of transformations determines whether the values of
$n_{eff}$ or $\boldsymbol{\varepsilon}$ and $\boldsymbol{\mu}$ are
high or low, positive or negative, or near zero. Therefore, it
needs to take all factors into consideration and select
appropriate values to realize the design by materials available.

\section{Discussion and Conclusion}
Before concluding this work it merits some further discussions.
First, to implement the above design, metamaterial is competent
without any technological difficulties \cite{Cui2010,Jiang2011}.
Now another simpler way is to further discretize the material
parameters and build by dielectrics, which has been proved
efficient to realize lossless and broadband transformation devices
\cite{Kundtz2010,Roberts2009,Yang2011,Li2008,Han2010,
Gharghi2011}. Second, according to the principle of equal optical
path the method can be generalized to realize arbitrary phase
conversion through media with arbitrary geometric shapes. The
coordinate transformation should guarantee that the needed OPLs
are satisfied. Last, we only considered the transformation in two
dimensions. Actually, the method is applicable to
three-dimensional problems and more new phenomena may be found.
For example, the method can be used to transform planar wavefront
of Gaussian beam into helical one of vortex beam carrying orbital
angular momentum \cite{Yao2011}.

In summary, we have proposed a planar phase transformer to achieve
the conversion between any two wavefronts. From the point of view
of OPL we present a general method of transformation wherein the
profile determined by the difference of the two wavefronts is
taken to be converted to a plane surface. For the mutual
conversion between planar and curved wavefronts, the method is
greatly simplified as the ITM in which it is the opposite shape of
the desired wavefront that is converted to a plane. Applying the
method, three types of phase transformation are investigated,
which can be used for wave deflector and flat lens, and are
further confirmed by numerical simulations. In addition, we find
that the transformation on phase can realize some important
properties such as phase reversal or compensation, focusing, and
expanding or compressing beams. The slab can be realized by
materials whose values of refractive index or
$\boldsymbol{\varepsilon}$ and $\boldsymbol{\mu}$ depend on the
choice of transformations. The method provides a new insight into
transformation optics and we expect it will be of practical
importance for designing compact optical devices.

\section{Acknowledgements}
This work was supported in part by the National Natural Science
Foundation of China (No.10847121, 10804029, 10904036, 61025024)
and the Growth Program for Young Teachers of Hunan University.

\end{document}